\documentclass[preprint, 3p, twocolumn]{elsarticle}
\usepackage{graphicx}
\usepackage{epstopdf}
\usepackage{amssymb}
\usepackage{lineno}
\usepackage{caption}
\journal{Journal of Magnetism and Magnetic Materials}

\begin{document}

\begin{frontmatter}

\title{Spin-flop phase transition in the orthorhombic antiferromagnetic topological semimetal Cu$_{0.95}$MnAs}

\author[ucla]{Eve Emmanouilidou }
\author[ucla]{Jinyu Liu }
\author[HFML]{David Graf }
\author[ornl]{Huibo Cao}
\author[ucla]{Ni Ni \corref{cor}}
\ead{nini@physics.ucla.edu}
\address[ucla]{Department of Physics and Astronomy, University of California, Los Angeles, California 90095, USA}
\address[HFML]{National High Magnetic Field Laboratory, 1800 E. Paul Dirac Drive, Tallahassee, FL 32310, USA
}
\address[ornl]{Quantum Condensed Matter Division, Oak Ridge National Laboratory, Oak Ridge, Tennessee 37831, USA}

\cortext[cor]{Corresponding author}

\begin{abstract}
The orthorhombic antiferromagnetic compound CuMnAs was recently predicted to be an antiferromagnetic Dirac semimetal if both R$_y$ gliding and S$_{2z}$ rotational symmetries are preserved in its magnetic ordered state. In our previous work on Cu$_{0.95}$MnAs and Cu$_{0.98}$Mn$_{0.96}$As, we showed that in their low temperature commensurate antiferromagnetic state the $\textit{b}$ axis is the magnetic easy axis, which breaks the S$_{2z}$ symmetry. As a result, while the existence of Dirac fermions is no longer protected, the polarized surface state makes this material potentially interesting for antiferromagnetic spintronics. In this paper, we report a detailed study of the anisotropic magnetic properties and magnetoresistance of Cu$_{0.95}$MnAs and Cu$_{0.98}$Mn$_{0.96}$As. Our study shows that in Cu$_{0.95}$MnAs the $b$ axis is the easy axis and the $c$ axis is the hard axis. Furthermore, it reveals that Cu$_{0.95}$MnAs features a spin-flop phase transition at high temperatures and low fields when the field is applied along the easy $b$ axis, resulting in canted antiferromagnetism. However, no metamagnetic transition is observed for Cu$_{0.98}$Mn$_{0.96}$As, indicating that the magnetic interactions in this system are very sensitive to Cu vacancies and Cu/Mn site mixing.
\end{abstract}


\end{frontmatter}

\section{Introduction}
Antiferromagnetic (AFM) materials have recently brought new excitement to the field of condensed matter physics due to their potential applications in spintronics, a field which studies the effect of the charge carrier spin in conduction. AFM systems lack a net magnetic moment although the individual atoms are magnetic, and this makes them ``invisible" to external magnetic fields, which originally led researchers to believe that they could not be used for practical applications  \cite{spintronics:review}. It was not until a few years ago that it was realized that antiferromagnets have many characteristics that make them suitable for spintronics; they are insensitive to magnetic field perturbations, do not generate stray fields, and have faster spin dynamics than ferromagnets since their resonant frequencies are higher  \cite{spintronics:review, proposal}. The prediction and subsequent discovery of the anomalous Hall effect \cite{Mn3Ir,Mn5Si3} and the spin Hall effect \cite{Ir20Mn80, CuAs, PtMn} in AFMs have also contributed to their recent popularity.

 Tetragonal CuMnAs has been studied for its potential applicability as an AFM spintronic material since 2013 \cite{cumnas:2013, cumnas:2015}. Recently, its orthorhombic polymorph was proposed to host Dirac fermions\cite{cumnas:prediction}. If the combination of inversion and time-reversal symmetries, $\mathcal{PT}$, is preserved, the existence of Dirac fermions is achieved when both R$_y$ gliding and S$_{2z}$ rotational symmetries are preserved in the magnetic state. The Dirac fermions in CuMnAs have also been predicted to be controlled by the spin-orbit torque reorientation of the N\'eel vector \cite{Dirac:Neel} .

In our previous paper, we first reported on the synthesis and magnetic structures of single crystalline Cu$_{0.95}$MnAs, which crystallizes in the $\textit{Pnma}$ space group with lattice parameters $\textit{a}$ = 6.5716(4) \AA, $\textit{b}$ = 3.8605(2) \AA, and $\textit{c}$ = 7.3047(4) \AA,  and Cu$_{0.98}$Mn$_{0.96}$As, with lattice parameters $\textit{a}$ = 6.5868(4) \AA, $\textit{b}$ = 3.8542(3) \AA, and $\textit{c}$ = 7.3015(5) \AA \citep{eve:2017}. Although an extra intermediate incommensurate AFM state exists in Cu$_{0.98}$Mn$_{0.96}$As, when both Cu$_{0.95}$MnAs and Cu$_{0.98}$Mn$_{0.96}$As are in their low-temperature commensurate AFM state, neutron diffraction measurements reveal that in the distorted Mn honeycomb sublattice the Mn spins order antiparallel to each of their nearest neighbors with the spins along the $\textit{b}$-axis. We concluded that this magnetic structure breaks the S$_{2z}$ symmetry, leading to the disappearance of Dirac fermions. This is consistent with a following study on CuMnAs single crystals based on the argument that the easy axis is in the $\textit{ab}$ plane \cite{cumnas:competition}. Our first-principles calculations show that this magnetic order can support spin-polarized states, a much sought after property for spintronics.

In this paper, we discuss the observation of a metamagnetic phase transition in Cu$_{0.95}$MnAs, and its absence in Cu$_{0.98}$Mn$_{0.96}$As, through a thorough study of magnetic susceptibility and magnetoresistance, and present a magnetic phase diagram for Cu$_{0.95}$MnAs.

\begin{figure}[h]
\centering\includegraphics[width=3.1in]{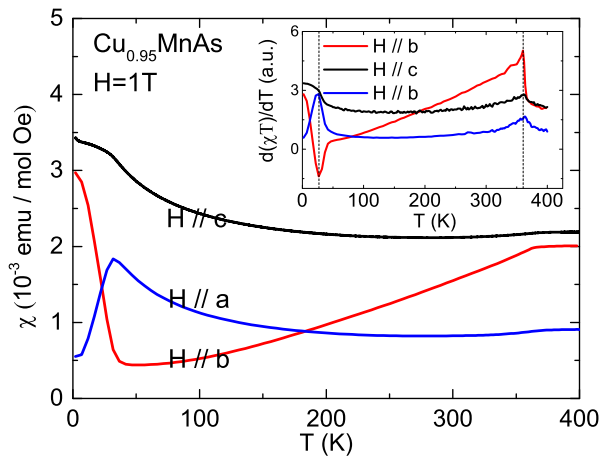}
\caption{Temperature dependence of the susceptibility ($\chi$) of Cu$_{0.95}$MnAs with a magnetic field of 1T, applied parallel to the $\textit{a}$,$\textit{b}$ and $\textit{c}$ crystallographic axes. Inset: The derivative of the quantity $\chi$T with respect to temperature.}
\end{figure}

\section{Experimental Methods}

CuMnAs single crystals were synthesized using the flux method. Chunks of Bi were used as the flux, with the recipes described in detail in Ref.\cite{eve:2017, klora}. Magnetic susceptibility data with the magnetic field applied parallel to the $\textit{a}$,$\textit{b}$ and $\textit{c}$ crystallographic axes were collected in a Quantum Design (QD) Magnetic Property Measurement System (MPMS). In order to acquire a large enough signal, four to five pieces were carefully aligned on a quartz holder that has minimal contribution to the magnetic signal. Electrical resistivity ($\rho_{xx}$) and magnetoresistance (MR) were measured in a QD Physical Property Measurement System (PPMS) under magnetic fields from -9 T to 9 T, and the National High Magnetic Field Lab (NHMFL) in Tallahassee, FL for fields up to 35T. The MR was calculated using equation MR = $\frac{\rho_{xx}(H) - \rho_{xx}(0)}{\rho_{xx}(0)}$, after $\rho_{xx}(H)$ had been symmetrized. An excitation current of a few mA was applied parallel to the $\textit{b}$ axis in all cases.
\section{Results and Discussion}

Fig. 1 shows the temperature dependence of the magnetic susceptibility $\chi$ of Cu$_{0.95}$MnAs under a magnetic field of 1 T, applied parallel to the $\textit{a}$,$\textit{b}$ and $\textit{c}$ axes. Cu$_{0.95}$MnAs undergoes a temperature induced second-order paramagnetic (PM) to AFM phase transition at 360 K and the effect of the transition is most pronounced when H//$\textit{b}$, as the susceptibility begins to decrease dramatically below this temperature. With H//$\textit{a}$ and H$//\textit{c}$, $\chi$ shows much smaller change across 360 K, which can be better seen in the $d(\chi T)/dT$ plot shown in the inset of Fig. 1. This suggests that the magnetic easy axis for Cu$_{0.95}$MnAs is the $\textit{b}$ axis, which is consistent with the neutron diffraction results\cite{eve:2017}. No Curie-Weiss behavior is observed up to 400 K.

\begin{figure}[h]
\centering\includegraphics[width=3.1in]{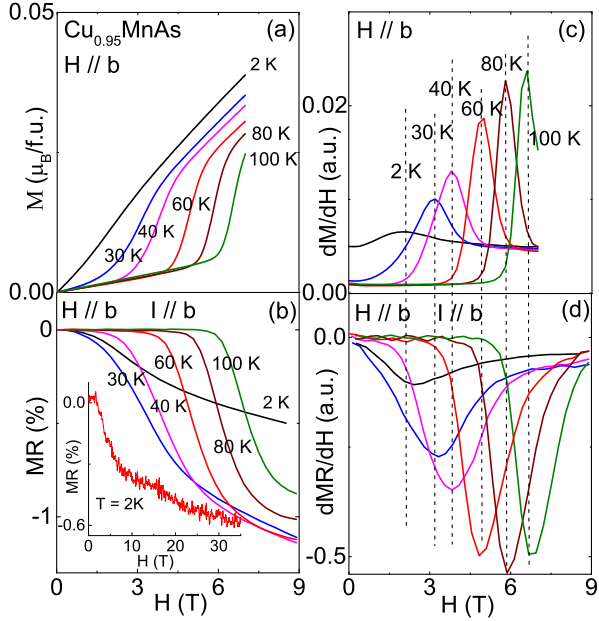}
\caption{(a): The magnetization, M, of Cu$_{0.95}$MnAs at several temperatures for H//$\textit{b}$. (b): The magnetoresistance, MR, of Cu$_{0.95}$MnAs at several temperatures, for H//$\textit{b}$ and I//$\textit{b}$. Inset: The MR for fields up to 35 T. (c)-(d): The derivatives of M and MR with respect to H. The dashed line shows the criterion to determine $H_{SF}$.}
\end{figure}

\begin{figure}[h]
\centering\includegraphics[width=3.1in]{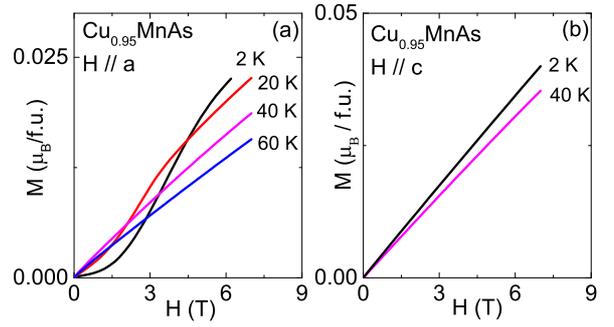}
\caption{(a): The field dependence of the magnetization of Cu$_{0.95}$MnAs for 2 K, 20 K, 40 K and 60 K with the field parallel to $\textit{a}$. (b): The field dependence of the magnetization of Cu$_{0.95}$MnAs for 2 K and 40 K with the field parallel to $\textit{c}$. }
\end{figure}

As the sample is further cooled, a second anomaly is observed around 27 K. In contrast to the high temperature transition, the susceptibility with H//$\textit{a}$ starts decreasing sharply, while increasing along the other two directions. For H//$\textit{b}$ the increase is dramatic; the susceptibility at 2 K is more than twice as large as that at 30 K. This transition can also be clearly seen in the inset of Fig. 1.

To investigate the nature of this low temperature transition, the isothermal magnetization and magnetoresistance of Cu$_{0.95}$MnAs with H//$\textit{b}$ for temperatures between 2 K and 100 K were measured and are shown in Fig.2. A metamagnetic phase transition, which broadly refers to the abrupt increase of the magnetization with applied magnetic field, can be seen very clearly as a sharp upturn in the isothermal magnetization data up to 100 K, as shown in Fig.2(a). The critical field H$_{SF}$, defined here as the maximum of $dM/dH$ (Fig.2(c)), moves to higher values as the temperature is increased, as shown in Fig.2(c). This behavior suggests a spin-flop transition, which was first experimentally observed in CuCl$_2$ $\cdot$  H$_2$O single crystals \cite{cucl2}, and has since also been observed in many other systems with magnetocrystalline anisotropy \cite{mntio3, ba3cu2o4cl2,cacoas, niwo4,nio,cuvo, laprcop,gd3co,fega2s4}. In a spin-flop transition with weak magnetocrystalline anisotropy, when a magnetic field is applied parallel to the magnetic easy axis of a material  and exceeds a critical value H$_{SF}$, the simple antiferromagnetic spins suddenly rotate into the spin-flop state where the spins try to align perpendicular to the magnetic field, and this results in a net moment along the easy axis \cite{theory}. Upon further increasing the field, the net moment grows until saturation. The spin-flop transition happens because the total energy of the system is a sum of the Zeeman energy of each magnetic sublattice and the magnetic anisotropy energy. At low fields the AFM configuration is the ground state, but above H$_{SF}$, it is the spin-flop state with spins almost normal to the field that minimizes the energy \cite{blundell}.

No hysteresis is observed in Fig.2(a), as is the case for most AFM materials with spin-flop transitions. For years the spin-flop transition had been considered to be a first-order transition, with the absence of the magnetic hysteresis attributed to low magnetic anisotropy, but its nature is now under debate\cite{Li}. No saturation is observed up to 7 T for Cu$_{0.95}$MnAs, which indicates that the system remains in the spin-flop state without saturation up to 7 T. This is consistent with the fact that the maximum magnetic moment at 2 K and 7 T is just 0.04 $\mu_B$/f.u., much smaller than the saturation moment of Mn$^{2+}$.

The spin-flop transition can also be clearly seen in the isothermal magnetoresistance data, shown in Fig.2(b). Associated with the transition is a sharp drop in the magnetoresistance to around -1\% at 9 T, suggesting the loss of spin scattering above H$_{SF}$, which is consistent with the spin-flop transition scenario. The inset of Fig.2(b) shows the magnetoresistance measured up to 35 T at 2 K. The magnetoresistance decreases linearly with field above 9 T, suggesting that the system remains in the spin-flop state up to 35 T without saturation.
The critical fields in both measurements are in very good agreement, as seen in Fig.2(c) and 2(d), which show the derivatives of the isothermal magnetization and the magnetoresistance with respect to field. The small mismatches can be ascribed to misalignment between the magnetic field direction and the magnetic easy axis.

\begin{figure}[h]
\centering\includegraphics[width=3.1in]{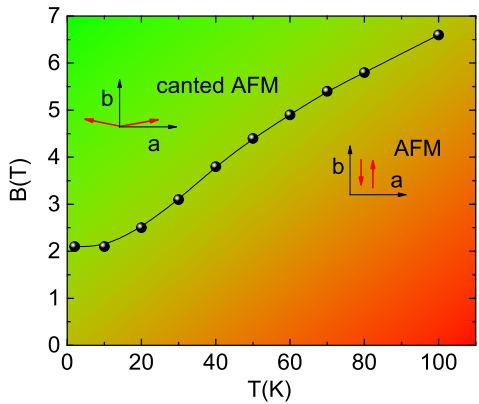}
\caption{The magnetic phase diagram of Cu$_{0.95}$MnAs with the field parallel to the $b$ axis. }
\end{figure}

To further investigate the magnetic anisotropy, the isothermal magnetization for Cu$_{0.95}$MnAs with H//$\textit{a}$ and H//$\textit{c}$ was measured and is shown in Fig. 3(a) and 3(b), respectively. With H//$\textit{a}$, the isothermal magnetization shows a metamagnetic transition below 30 K, while linearly increasing with field above 30 K. This is in sharp contrast to what we observe when $H // b$,  where the metamagnetic transition can be seen even at 100 K. When the field is parallel to the $\textit{c}$ direction, the magnetization remains linear with field at both 2 K and 40 K up to 7 T, showing no sign of a metamagnetic transition. This behavior suggests that the $\textit{b}$ axis is the easy axis and the $\textit{c}$ axis is the hard axis.

Based on what we have observed in Figs. 1-3, we suggest that this spin-flop transition results in a canted antiferromagnetic state when a small magnetic field is applied, as shown in the inset of Fig. 4. A proposed magnetic phase diagram for Cu$_{0.95}$MnAs, based on our results in magnetic susceptibility, magnetization and magnetoresistance with the field parallel to the easy axis is shown in Fig. 4. The $H_{SF}$ at each temperature is determined using the criterion shown as the dashed line in Fig. 2(c). At high temperatures and below H$_{SF}$, the Mn spins are oriented along the $\textit{b}$ axis. Above H$_{SF}$, the magnetic structure consists of the Mn spins that have now flipped and form a small canting angle with the $\textit{a}$ axis that causes the magnetization to have a small component along $b$.

\begin{figure}[h]
\centering\includegraphics[width=3.1in]{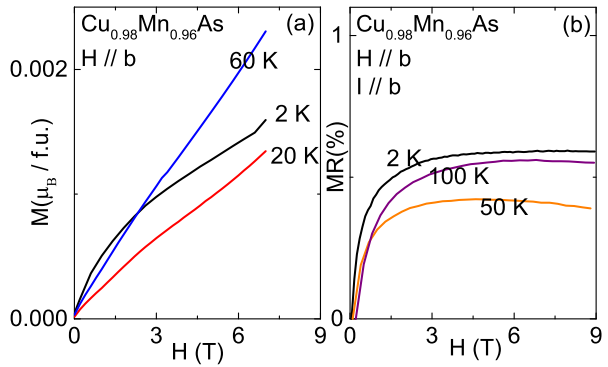}
\caption{(a)-(b): The isothermal magnetization (a) and magnetoresistance (b) of  Cu$_{0.98}$Mn$_{0.96}$As for certain temperature values. The magnetic field was applied parallel to the $b$ axis. }
\end{figure}

In our previous study, we showed that in orthorhombic CuMnAs, the magnetism is very sensitive to the Cu vacancies and Cu/Mn site mixing \cite{eve:2017}. While Cu$_{0.95}$MnAs is a commensurate antiferromagnet below 360 K, Cu$_{0.98}$Mn$_{0.96}$As enters an incommensurate antiferromagnetic state at 320 K and then a commensurate antiferromagnetic state below 230 K. To further examine the effect of Cu vacancies and Cu/Mn site mixing on the spin-flop phase transition, we performed isothermal magnetization and magnetoresistance measurements on Cu$_{0.98}$Mn$_{0.96}$As. The data are summarized in Figs. 5(a) and (b), respectively. No sign of a spin-flop transition is observed. The field-dependent magnetization evolves from a convex shape at 2 K to linear behavior at 60 K, with no sharp upturn. The MR is positive and quickly plateaus, reaching a maximum value of ~0.5$\%$. Therefore, a few \% of Cu vacancies or mixing with Mn atoms can destroy the spin-flop transition, suggesting a high sensitivity of the magnetism to the defects in orthorhombic CuMnAs.

\section{Conclusion}
In conclusion, we have studied the magnetization and magnetoresistance of Cu$_{0.95}$MnAs and Cu$_{0.98}$Mn$_{0.96}$As single crystals. Cu$_{0.95}$MnAs, as evidenced by both types of measurements, undergoes a metamagnetic spin-flop transition at high temperatures and low fields with the field along the easy axis $b$, leading to a canted antiferromagentic state with a small net moment along the $b$ axis. On the other hand, in Cu$_{0.98}$Mn$_{0.96}$As these transitions are absent, suggesting a dramatic effect of Cu vacancies or mixing with Mn atoms on the magnetism in the orthorhombic CuMnAs compound.

\section*{Acknowledgments}
Work at UCLA was supported by the U.S. Department of Energy (DOE), Office of Science, Office of Basic Energy Sciences under Award Number DE-SC0011978 and the NSF-MRI grant 1625776. The National High Magnetic Field Laboratory is supported by the National Science Foundation through NSF/DMR-1644779 and the State of Florida.

\end{document}